\documentclass[review]{elsarticle}
\makeatletter
\def\ps@pprintTitle{%
	\let\@oddhead\@empty
	\let\@evenhead\@empty
	\def\@oddfoot{\centerline{\thepage}}%
	\let\@evenfoot\@oddfoot}
\makeatother

\usepackage{lineno,hyperref}
\usepackage{threeparttable}
\usepackage{pdflscape}
\usepackage{float}
\usepackage{color}

\usepackage{comment}
\newcommand{\comm}[1]{}

\usepackage{xr}

\modulolinenumbers[5]
\journal{ }

\begin{document}

\begin{frontmatter}

\title{New ZrB$_2$ polymorphs: First-principles calculations}

\author[]{Marcin Ma\'zdziarz\corref{mycorrespondingauthor}}
\ead{mmazdz@ippt.pan.pl}

\author{ Tomasz Mo\'scicki}

\address{Institute of Fundamental Technological Research Polish Academy of Sciences,
Pawi\'nskiego 5B, 02-106 Warsaw, Poland}
\cortext[mycorrespondingauthor]{Corresponding author}





\begin{abstract}
Two new hypothetical zirconium diboride (ZrB$_2$) polymorphs: (\textit{hP6}-P6$_3$/mmc-space group, no.194) and (\textit{oP6}-Pmmn-space group, no.59),  were thoroughly studied under the first-principles density functional theory calculations from the structural, mechanical and thermodynamic properties point of view. As opposed to the known phase (\textit{hP3}-P6/mmm-space group, no.191) are not brittle. Knowledge about these new phases can be very useful when doping metal borides with zirconium.
\end{abstract}

\begin{keyword}
Zirconium diboride\sep
\textit{Ab initio} calculations\sep
Mechanical properties\sep
Elastic properties\sep
Phonons
\end{keyword}

\end{frontmatter}


\section{Introduction}

During latest years the transition metal borides due to combination of their outstanding physical properties such as electric and thermal conductivity comparable with metals, low compressibility, high shear strength, and exceptionally high hardness have attracted attention among materials researchers \cite{Akopov2017, Akopov2018}. Even in the form of thin films they poses extraordinary properties such as very high hardness with increased flexibility, great thermal properties and very good corrosion and wear resistance \cite{Fuger2019, MIRZAEI2020125274, MOSCICKI2020125689}.  Among borides, zirconium diboride (ZrB$_2$) deserves special attention. ZrB$_2$ with melting temperature 3245$^{\circ}$C \cite{Fahrenholtz2007} is a member of a family of materials known as ultra high-temperature ceramics (UHTCs). In addition to high melting temperatures, ZrB$_2$ has a unique combination of chemical stability, high electrical and thermal conductivities, resistance to erosion and corrosion that makes its suitable for the extreme chemical and thermal environments associated with for example hypersonic flight, atmospheric re-entry, and rocket propulsion \cite{Opeka2004OxidationbasedMS}.
According to the Zr–B phase diagram \cite{CHEN2009209, 2008MB200809} there are three phases, namely, ZrB, ZrB$_2$, and ZrB$_{12}$, which have been reported and widely studied for this system. Theoretical investigations show that ZrB can create different crystallographic structures. The basic phase is NaCl-type face-centered cubic ZrB (\textit{Fm-3m}-space group, no.225) with lattice constant a = 4.900\AA \cite{LI2010814}. Furthermore, ZrB also can crystallize in FeB - type structure with a primitive orthorhombic (Pnma) crystal structure \cite{2008MB200809, HUANG20156831}, CrB-type orthorhombic structure with \textit{Cmcm}-space group \cite{HUANG20156831, XU2013105} and hexagonal  \textit{Pmmm} \cite{Li2019}. 
The literature review shows, ZrB$_{12}$ is stable in only one structure type. The cubic LuB$_{12}$ structure (\textit{Fm3m}-space group, no.225) with lattice constant a = 7.4085\AA{} was studied theoretically and experimentally for example in \cite{Jger2006}. They compared electric-field gradient measurements at the B sites and first-principles calculations in order to analyse the chemical bonding properties. Both, experimental and theoretical results were in good agreement. Whereas in \cite{Pan2020} the mechanical properties of ZrB$_{12}$ were calculated. The Vickers hardness of zirconium boride was 32.9\,GPa, which is in good agreement with the other theoretical result \cite{Jger2006, JLi2015, CHEN20166624}. 

In contrast to ZrB$_{12}$, depending on the calculation method, different hardness values have been determined for zirconium diboride ZrB$_2$. The calculated \cite{ZHANG2008411} values of hardness range from 12.82\,GPa to 55.53\,GPa, whereas experimentally measured hardness reaches 23$\pm$0.9\,GPa \cite{Fahrenholtz2007, Chamberlain2008}. All mentioned structures were assigned as ZrB$_2$ with the crystal hexagonal structure of AlB$_2$-type with the \textit{P6/mmm}-space group, no.191. Such large differences in the values of the analysed properties may, however, come not only from differences in calculation methods but also from the possibility of existence of other stable forms of ZrB$_2$ which may form nanocomposite of different polymorphs of ZrB$_2$. A similar conclusion about the possibility of existence of other ZrB$_2$ crystal types can be drawn on the basis of other studies on possible forms of transition metal diborides, e.g. WB$_2$ \cite{MAZDZIARZ201692, Cheng2014} or ReB$_2$ \cite{Marco2016, Mazdziarz2016}. For comparison in \cite{Cheng2014} the Authors proposed six different phases of WB$_2$. In the case of ZrB$_2$ is hard to find such study.

It should be also noted that in addition to the phases appearing in the Zr-B equilibrium diagram \cite{CHEN2009209, TOGO20151}, other zirconium and boron compounds were theoretically determined. There are hypothetical Zr-B phases such as: Zr$_3$B$_4$, Zr$_2$B$_3$, Zr$_3$B$_2$ \cite{JLi2015}, ZrB$_3$ \cite{ma9080703}, ZrB$_4$ \cite{Zhang2013b} and ZrB$_6$ \cite{Li2019}. All polymorphs are both mechanically and dynamically stable but have not been confirmed experimentally yet.
In this work structural, mechanical and thermodynamic properties of stable ZrB$_2$ polymorphs from density functional calculations will be studied.

\section{Computational methods}
\label{sec:Cm}

First-principle calculations based on density functional theory (DFT) \cite{DFT-HK, DFT-KS} within the pseudopotential, plane-wave approximation (PP-PW) implemented in ABINIT \cite{GONZE2016106, GONZE2020107042} code were performed in this work. 
Projector augmented-wave formulation (PAW) pseudopotentials \cite{Martin2019} were used to describe the interactions of ionic core and non-valence electrons. 

To enhance the confidence of the calculations as an exchange-correlation (XC) functional three approximations were used: local density approximation (LDA) \cite{BlochBemerkungZE1929,Perdew1992}, classical Perdew-Burke-Ernzerhof (PBE) generalized gradient approximation (GGA) \cite{Perdew1996} and modified Perdew-Burke-Ernzerhof GGA for solids (PBEsol) \cite{Perdew2008}. There is a strong view that the PBEsol is the overall best performing XC functional for identifying the structure and elastic properties \cite{Rasander2015, MAZDZIARZ20177, Mazdziarz2018}.

Used PAW pseudopotentials for LDA and PBE XC functionals were taken from PseudoDojo project \cite{JOLLET20141246}. Projector augmented wave method (PAW) pseudopotentials for PBEsol exchange-correlation functional \cite{Perdew2008} were generated using ATOMPAW software \cite{HOLZWARTH2001329} and a library of exchange-correlation functionals for density functional theory LibXC \cite{LEHTOLA20181}.

All calculations were made by tuning the precision of the calculations by automatically setting the variables at \textit{accuracy} level 4. The \textit{cut-off} energy consistent with (PAW) pseudopotentials of the plane-wave basis set was 15\,Ha  with the $4d^25s^2$ valence electrons for Zr and $2s^22p^1$ valence electrons for B. K-PoinTs grid was generated with \textit{kptrlen}=30.0. Metallic occupation of levels with Fermi-Dirac smearing occupation scheme and \textit{tsmear} (Ha) = 0.02 was used in all ABINIT calculations.

\subsection{Optimization of structures}
\label{ssec:Oos}
As mentioned earlier tungsten diboride \cite{MAZDZIARZ201692} and rhenium diboride \cite{Mazdziarz2016} crystallize in many various space groups. Searching for new structures of ZrB$_2$ we started with basic cells of \textit{hP6}-P6$_3$/mmc-WB$_2$ and \textit{oP6}-Pmmn-WB$_2$ and replaced tungsten atoms by zirconium atoms. Then all structures were relaxed by using the Broyden-Fletcher-Goldfarb-Shanno minimization scheme (BFGS) with full optimization of cell geometry and atomic coordinates. Maximal stress tolerance (GPa) was set to 1$\times$10$^{-4}$.
\subsection{Formation enthalpy and Cohesive energy}
\label{ssec:Ce}
The formation enthalpy and cohesive energy were determined as follows \cite{Qi20145843, Mazdziarz2016}:
\begin{eqnarray}
\triangle_f H({ZrB}_{2})={E}_{coh} ({ZrB}_{2}) - {E}_{coh} ({Zr}) - 2{E}_{coh} ({B}),
\label{eqn:Enthalpy}
\end{eqnarray}

\begin{eqnarray}
  {E}_{coh} ({ZrB}_{2})={E}_{total} ({ZrB}_{2}) - {E}_{iso} ({Zr}) - 2{E}_{iso} ({B}),
    \label{eqn:Ecoh}
\end{eqnarray}

where $\triangle_f H({ZrB}_{2})$ is the  formation enthalpy of the {ZrB}$_2$; $E_{coh}(ZrB_2)$ is the cohesive energy of the {ZrB}$_2$; $E_{coh}(Zr)$ is the cohesive energy of $Zr$; $E_{coh}(B)$ is the cohesive energy of $B$; $E_{tot}(ZrB_2)$ is the total energy of the {ZrB}$_2$; $E_{iso}(Zr)$ is the total energy of a $Zr$ atom and $E_{iso}(B)$ is the total energy of a $B$ atom.

\subsection{Mechanical properties calculations}
\label{ssec:EcHc}
The theoretical ground state elastic constants $C_{ij}$ of all structures were established with the metric tensor formulation of strain in density functional perturbation theory (DFPT) \cite{Hamann2005}. Isotropised bulk modulus \textbf{$B$}, shear modulus \textbf{$G$}, Young's modulus $E$ and Poisson's ratio $\nu$ were estimated by means of a \textit{Voigt–Reuss–Hill} average \cite{Hill_1952, Mazdziarz2015}.

In order to verify the elastic stability of all the structures positive definiteness of the stiffness tensor was checked \cite{Grimvall2012} by calculating Kelvin moduli, i.e. eigenvalues of stiffness tensor represented in \textit{second-rank tensor} notation \cite{Mazdziarz2019}. 

\comm{Generally to analyse elastic stability of the solid subjected to an external load Born stability condition \cite{Born1940,Misra1940} must be modified
\cite{Milstein1979,Milstein1998,Wang1995,Djohari2007}. The stability of the lattice at long-wavelengths unfortunately do not imply short-wavelength phonons stability \cite{Grimvall2012}.
\cite{Hill_1952} "The Elastic Behaviour of a Crystalline Aggregate"
\cite{Mazdziarz2019} proper elastic stability
\cite{Mazdziarz2018} i \cite{MAZDZIARZ20177}
}

Hardness of  ZrB$_2$ polymorphs in the present paper was calculated with the use of semi-empirical relation proposed in \cite{Tian201293}. The equation is defined as follow:
\begin{eqnarray}
  H_v=0.92 (G/B)^{1.137}G^{0.708}.
    \label{eqn:Hardness}
\end{eqnarray}
\textit{G/B} ratio appearing in the above formula named Pugh's modulus ratio \cite{Pugh1954} is commonly used as a universal ductile-to-brittle criterion.

\subsection{Phonon and Thermodynamic properties calculations}
\label{ssec:Pc}
To calculate phonons, density functional perturbation theory (DFPT) was utilised \cite{GONZE2016106, GONZE2020107042}. The phonon dispersion curves \cite{HINUMA2017140} of the analysed structures were then used to determine their dynamical stability \cite{Grimvall2012,Petr2012} complementary to elastic stability. 
Acoustic Debye temperature was calculated from the phonon densities of states (DOS).

\label{ssec:Tp}
Using the calculated phonons under the harmonic approximation, i.e. in the range up to Debye temperature, thermal quantities: {phonon internal energy}, {free energy}, {entropy}, {constant volume heat capacity} as a function of the temperature were determined \cite{TOGO20151, BOTTIN2020107301}.

\section{Results}
\label{sec:Res}
Using the approach described in Sec.\ref{ssec:Oos} the first step in our calculations was the geometry optimization of the two new hypothetical zirconium diboride (ZrB$_2$) polymorphs: (\textit{hP6}-P6$_3$/mmc-space group, no.194) and (\textit{oP6}-Pmmn-space group, no.59) and the formerly known polymorph (\textit{hP3}-P6/mmm-space group, no.191).
\subsection{Structural properties }
\label{ssec:RSP}
\begin{figure}[H] 
	\centering
	\begin{tabular}{cc}
		\includegraphics[width=0.40\linewidth]{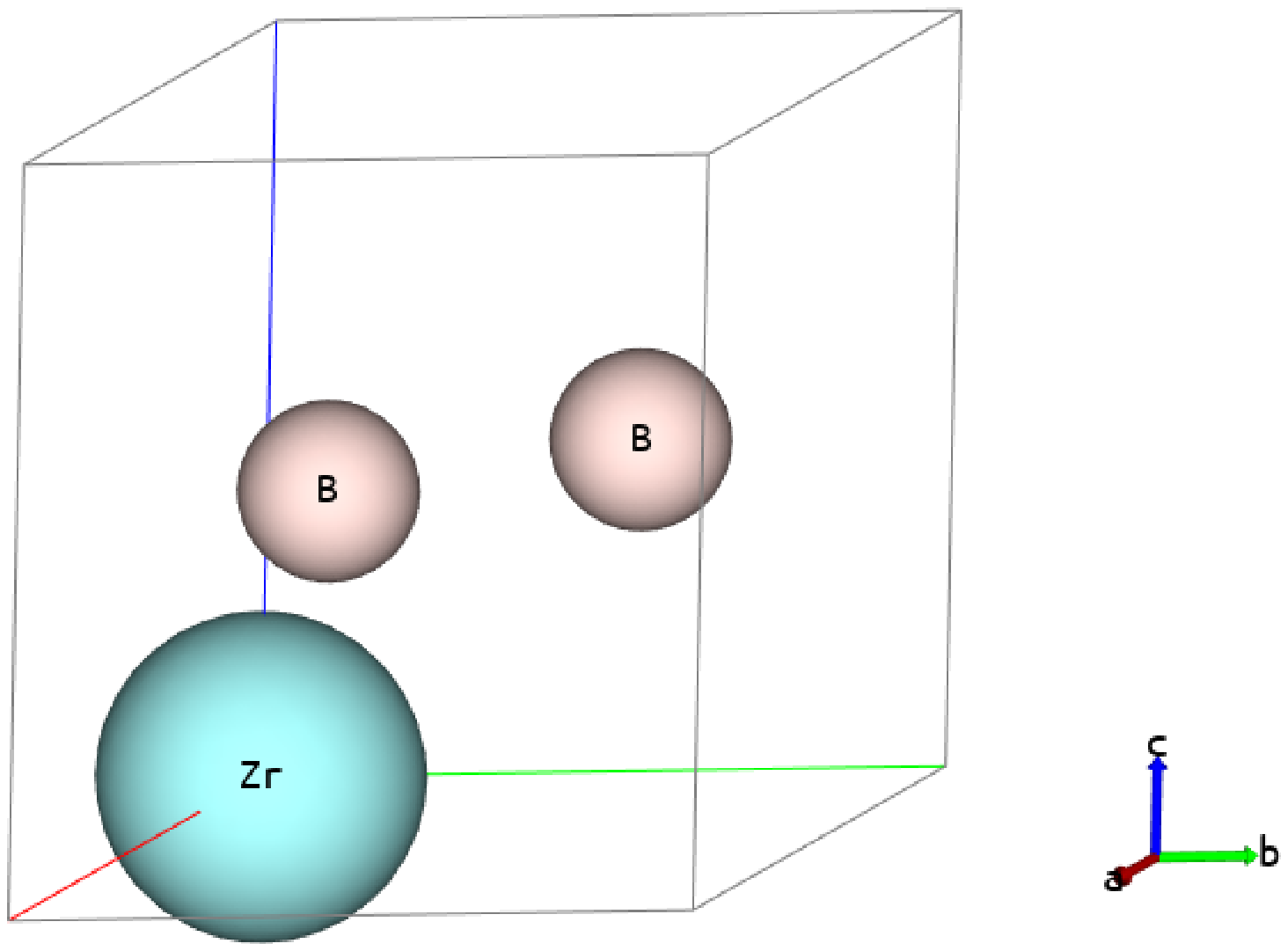} &
		\includegraphics[width=0.50\linewidth]{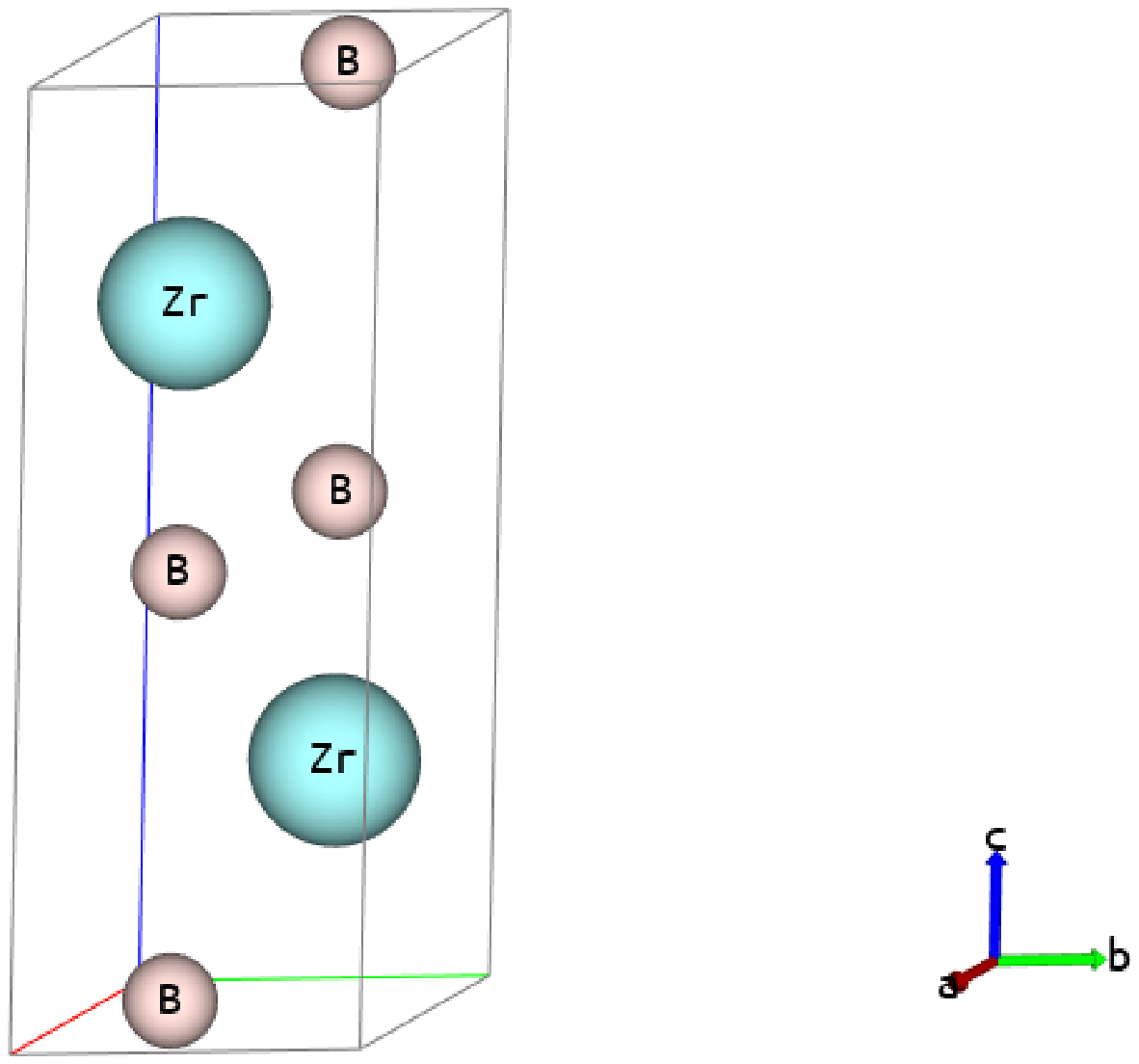} \\
		{\bf (a)} & {\bf (b)}\\
		\includegraphics[width=0.45\linewidth]{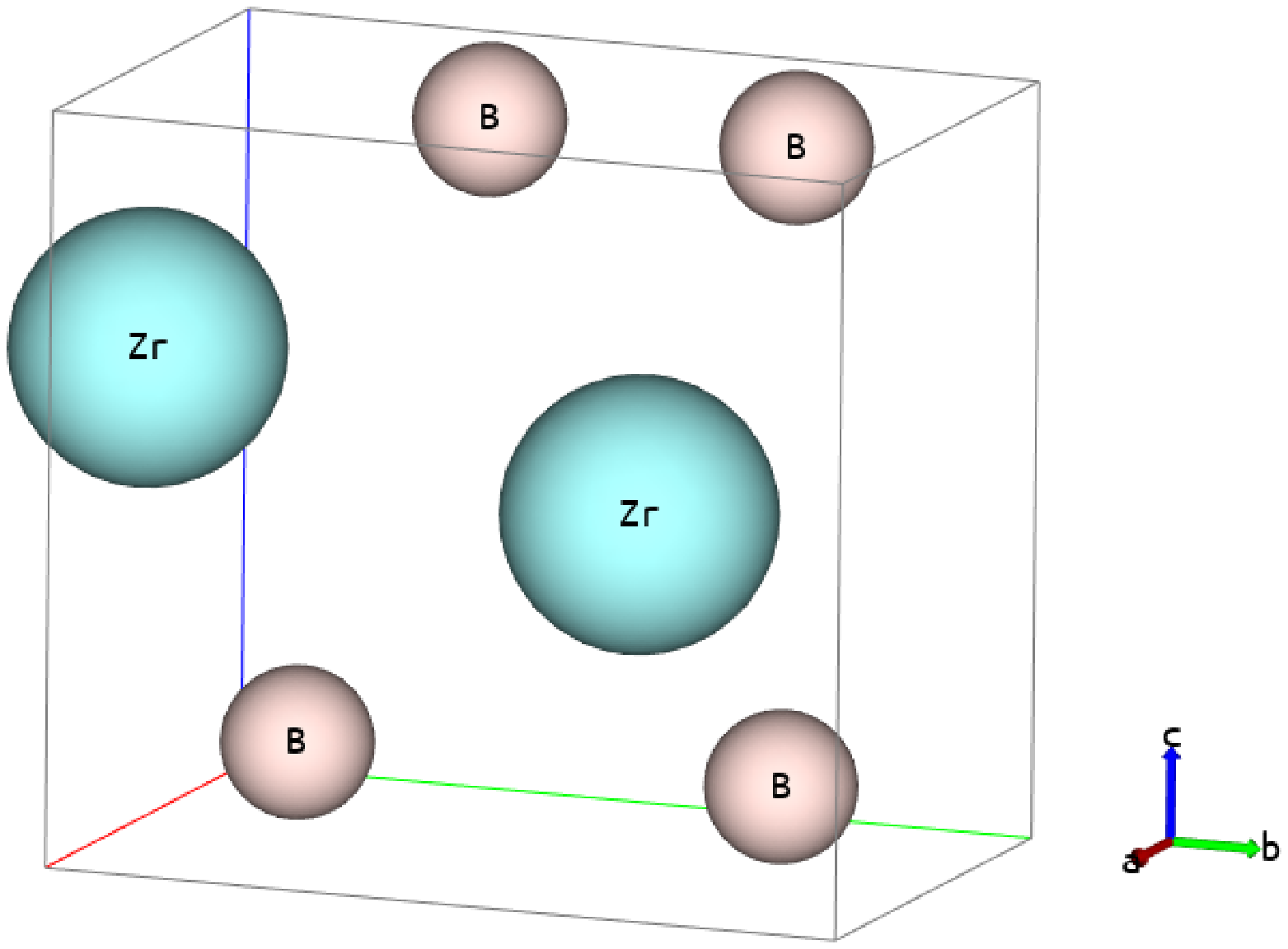} \\
		{\bf (c)}
	\end{tabular}
	\caption{ZB$_2$-Basic cells: a) \textit{hP3}-P6/mmm, b) \textit{hP6}-P6$_3$/mmc, c) \textit{oP6}-Pmmn }
	\label{fig:ZrB2Basiccells}
\end{figure}

The basic cells for all three analysed polymorphs are depicted in Fig.\ref{fig:ZrB2Basiccells}, whereas the crystallographic data, calculated with three different exchange-correlation (XC) functionals (LDA, PBE and PBEsol), are stored in Crystallographic Information Files (CIF) in {\ref{app:Appendix}. Supplementary data}.

\comm{Determined lattice parameters for known \textit{hP3}-P6/mmm, i.e. (a=3.135$\div$3.173\,{\AA} and c=3.477$\div$3.527\,{\AA}), are comparable to those of other authors, see Tab.\ref{tab:ZrB2Results}, a similar trend can be seen for calculated formation enthalpy (-$\triangle_f H$=1.145$\div$1.078\,eV/Atom) and cohesive energy (-$E_c$=8.769$\div$8.072\,eV/Atom).}

Determined lattice parameters, formation enthalpy and cohesive energy for known \textit{hP3}-P6/mmm polymorph, Fig.\ref{fig:ZrB2Basiccells} a), are comparable to those of other authors \cite{ZHANG2008411}, see Tab.\ref{tab:ZrB2Results}. We treat this as a verification of the correctness of the methodology used.

The first new hypothetical phase \textit{hP6}-P6$_3$/mmc, Fig.\ref{fig:ZrB2Basiccells} b), also crystallises in the hexagonal system but has 6 atoms in the cell, whereas the second new hypothetical phase \textit{oP6}-Pmmn, Fig.\ref{fig:ZrB2Basiccells} c), crystallises in the orthorhombic system and also has 6 atoms in the cell. There is a little sense to compare lattice parameters for phases in different systems, but it is worth to compare the formation enthalpy and the cohesive energy. It can be seen that the formation enthalpy $\triangle_f H$ for new phases is significantly higher than for known \textit{hP3}-P6/mmm and comparable between the new phases, see Tab.\ref{tab:ZrB2Results}. The calculated cohesive energy $E_c$ is a little higher than for known \textit{hP3}-P6/mmm and again comparable between the two new phases. These two facts suggest that the new phases are comparably thermodynamically stable but less stable than the known \textit{hP3}-P6/mmm phase.


\comm{\begin{figure}[H]
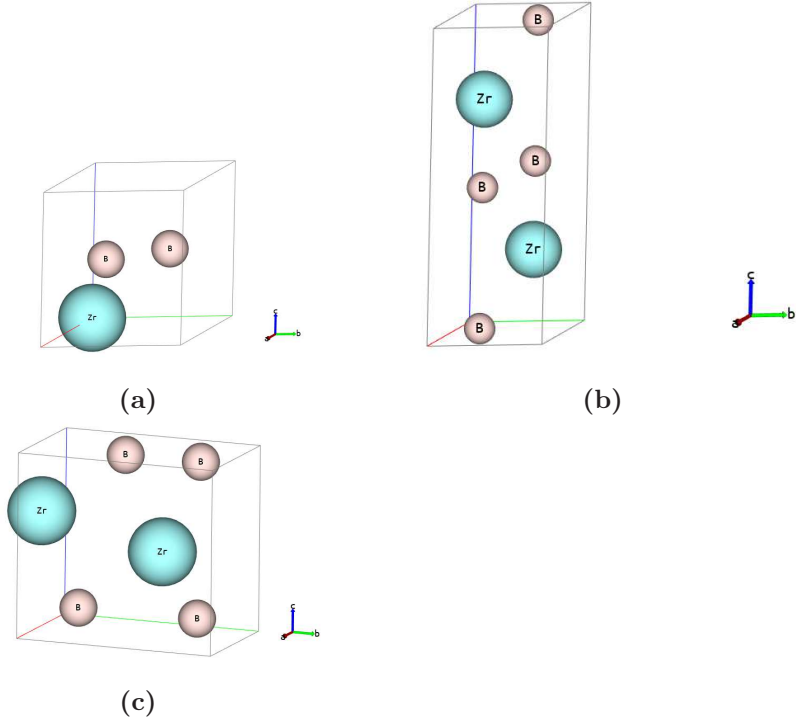
 
	\centering
	\begin{tabular}{c}
		\includegraphics[width=1.10\linewidth]{ZrB2_191.eps} \\
		{\bf (a)}\\
		\includegraphics[width=1.20\linewidth]{ZrB2_194.eps} \\
		{\bf (b)}\\
		\includegraphics[width=1.20\linewidth]{ZrB2_59.eps} \\
		{\bf (c)}
	\end{tabular}
	\caption{ZB$_2$-Basic cells: a) \textit{hP3}-P6/mmm, b) \textit{hP6}-P6$_3$/mmc, c) \textit{oP6}-Pmmn }
	\label{fig:ZrB2Basiccell}
\end{figure} }

\subsection{Mechanical properties }
\label{ssec:RMP}
Computed elastic constants, Kelvin moduli, isotropised bulk, shear and Young's modulus, Poisson's ratio, \textit{G/B} Pugh's modulus ratio, Debye temperature and estimated hardness of all analysed zirconium diboride structures are listed in Tab.\ref{tab:ZrB2Results}. For known \textit{hP3}-P6/mmm phase, Fig.\ref{fig:ZrB2Basiccells} a), these quantities are comparable to those of other authors \cite{ZHANG2008411}, what is a further verification of the validity of the methodology used.
Analysing the data received, it can be concluded that the new phases have lower mechanical parameters than the known \textit{hP3}-P6/mmm phase, except for the Poisson's ratio. Both new phases have similar isotropised bulk modulus of about 170\,GPa, but the shear modulus for the \textit{hP6}-P6$_3$/mmc phase is much lower and is only about 40\,GPa.  The consequence of this is that the hardness for  \textit{hP6}-P6$_3$/mmc phase is only 2\,GPa, while for \textit{oP6}-Pmmn phase it is about 14\,GPa. 
A high \textit{G/B} Pugh's modulus ratio would correspond to a more brittle character of material than ductile. The critical value, separating ductile from brittle materials, is approximately 0.571 \cite{Pugh1954, Rached201393}. It can be seen that known \textit{hP3}-P6/mmm phase is brittle, \textit{hP6}-P6$_3$/mmc phase is ductile and \textit{oP6}-Pmmn phase is somehow between brittle and ductile. 

All analysed structures have positive definite stiffness tensor, positive Kelvin moduli, i.e. eigenvalues of stiffness tensor represented in \textit{second-rank tensor} notation, so there are mechanically stable, see Tab.\ref{tab:ZrB2Results}. 

\begin{table}[H]
	\begin{threeparttable}[b]
		\centering
		\caption{Lattice parameters\,(\AA), formation enthalpy $\triangle_f H$\,(eV/Atom), cohesive energy $E_c$\,(eV/Atom), elastic constants  $C_{ij}$\,(GPa), Kelvin moduli $K_i$\,(GPa), bulk modulus $B$\,(GPa), shear modulus $G$\,(GPa), Young's modulus $E$\,(GPa), Poisson's ratio $\nu$, $G/B$ Pugh's modulus ratio, Debye temperature $\Theta_D$\,(K), hardness $H_v$\,(GPa) of ZrB$_2$ phases: ZrB$_2$(\textit{hP3}-P6/mmm-space group, no.191, \textit{hP6}-P6$_3$/mmc-space group, no.194, \textit{oP6}-P6/mmm-space group, no.59). Experimental and calculated data for \textit{hP3}-P6/mmm phase are taken from \cite{ZHANG2008411}.
			\label{tab:ZrB2Results}}
		\renewcommand{\arraystretch}{1.5}
		\tiny
		\begin{tabular}{|c c c c c c c c c c c|}
			\hline Phase &\multicolumn{4}{ c }{\textit{hP3}-P6/mmm-No.191} &\multicolumn{3}{ c }{\textit{hP6}-P6$_3$/mmc-No.194} & \multicolumn{3}{ c| }{\textit{oP6}-Pmmn-No.59} \\ 
			\hline Source  &  Exp./Calc. & LDA & PBE & PBEsol &  LDA  & PBE &  PBEsol & LDA &  PBE  & PBEsol \\
			\hline $a$ & 3.165$\div$3.169 &3.135 &3.173 &3.156 & 3.025 &3.076 &3.050 & 3.057 & 3.100 & 3.071\\
			$b$ &  &  &  &  &  &  &  &4.931 &5.029 &4.981\\
			$c$ &3.523$\div$3.547  &3.477 &3.527 &3.495 &8.515 &8.624 &8.565 &4.541 &4.604 &4.578 \\
			$-\triangle_f H$ &0.985$\div$1.099 &1.145 &1.078 &1.141 &4.023 &3.691 &3.853 &3.795 &3.654 &3.813\\
			$-E_c$ &8.648&8.769 &8.072 &8.411 &7.834 &7.187 &7.488 &7.187 &7.150 &7.448 \\
			$C_{11}$ &581 &618 &591 &597 &224 &214 &217 &333 &325 &336 \\
			$C_{22}$ &581 &618 &591 &597 &224 &214 &217 &331 &316 &334 \\
			$C_{33}$ &445 &477 &481 &456 &495 &447 &479 &436 &380 &439 \\
			$C_{44}$ &240 &278 &253 &269 &81  &73  &80  &136 &134 &144 \\
			$C_{55}$ &240 &278 &253 &269 &81  &73  &80  &43  &48  &35 \\
			$C_{66}$ &263 &283 &272 &274 &1   &9   &2   &122 &145 &126 \\
			$C_{12}$ &55  &52  &47  &49  &222 &196 &213 &111 &80  &109 \\
			$C_{13}$ &121 &135 &105 &126 &70  &63  &69  &97  &70  &88 \\
			$C_{23}$ &121 &135 &105 &126 &70  &63  &69  &99  &85  &92 \\
			$K_{I}$ & &787&727&753&572&519&555&578&500&569\\
			$K_{II}$ & &566&544&548&162&146&160&301&290&314\\
			$K_{III}$ & &566&544&548&162&146&160&272&283&288\\
			$K_{IV}$ & &556&506&538&2&18&4&244&268&252 \\
			$K_{V}$ & &556&506&538&2&18&4&221&238&226 \\
			$K_{VI}$ & &360&392&349&369&338&354&86&96&70 \\
			$B$ &220$\div$245&262 &242 &250 &184 &168 &178 &189 &165 &187 \\
			$G$ &225$\div$243&256 &247 &248 &37  &44  &37  &102 &108 &100 \\
			$E$ &502$\div$554&580 &554 &560 &104 &121 &105 &260 &267 &256 \\
			$\nu$&0.109$\div$0.13 &0.13&0.118&0.126 &0.406&0.38&0.402 &0.271 &0.231 &0.272 \\
			$G/B$&1.023$\div$0.992&0.981&1.024&0.995&0.200&0.261 &0.21&0.541 &0.655 &0.539 \\
			$\Theta_D$&910&1007&973&971&794&754&779&787&752&774 \\
			$H_v$&23$\div$55 &46 &47   &45   &2  &3 &2&12 &16 &12 \\
			\hline 
		\end{tabular}
	\end{threeparttable}
\end{table}

\subsection{Phonon and Thermodynamic properties}
\begin{figure}[H] 
	\centering
	\begin{tabular}{c}
		\includegraphics[width=0.98\linewidth]{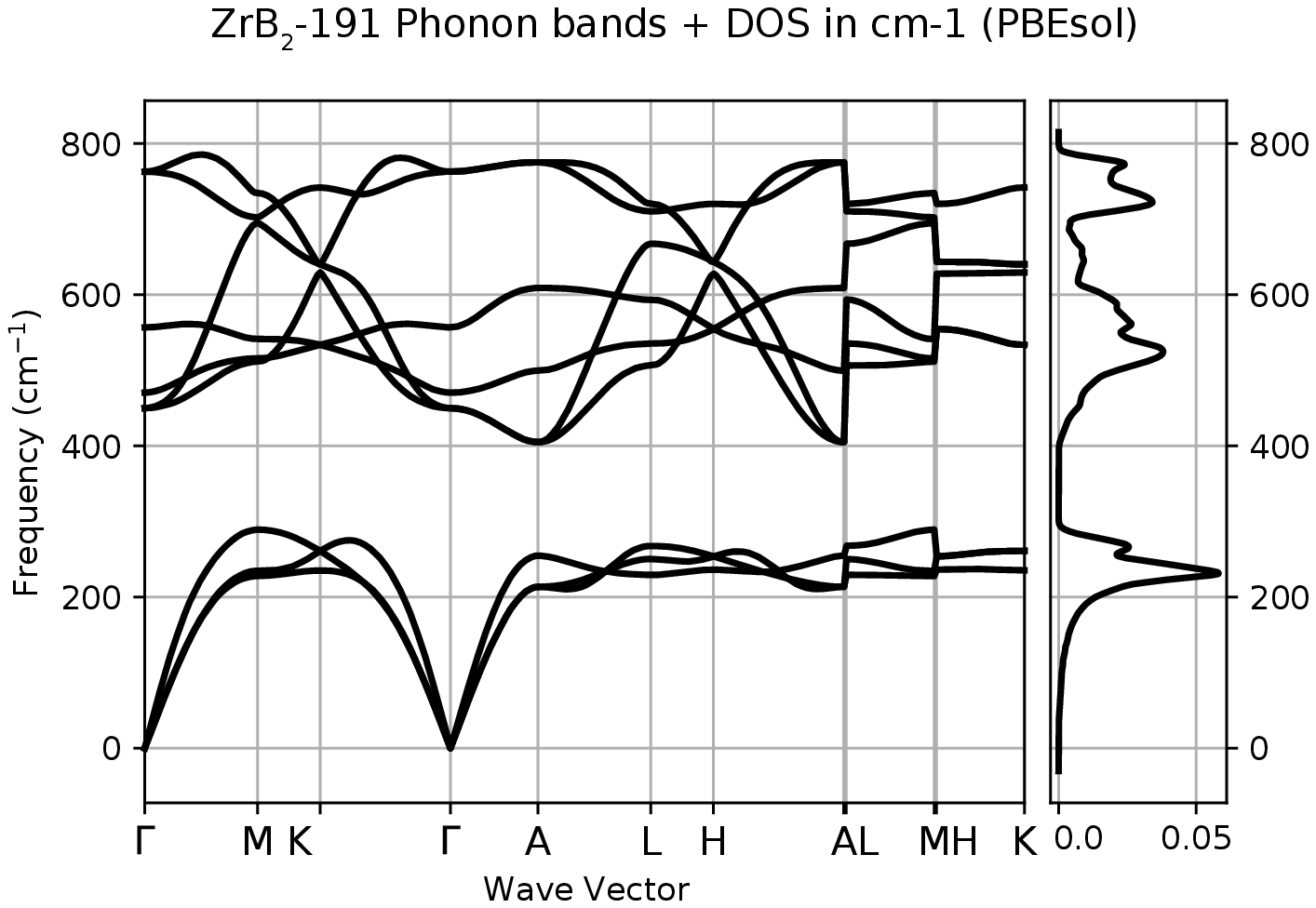} 
	\end{tabular}
	\caption{ZrB$_2$(\textit{hP3}-P6/mmm-space group, no.191)-Phonon band structure and DOS.}
	\label{fig:191PhononDispersion}
\end{figure}

\begin{figure}[H] 
	\centering
	\begin{tabular}{c}
		\includegraphics[width=0.98\linewidth]{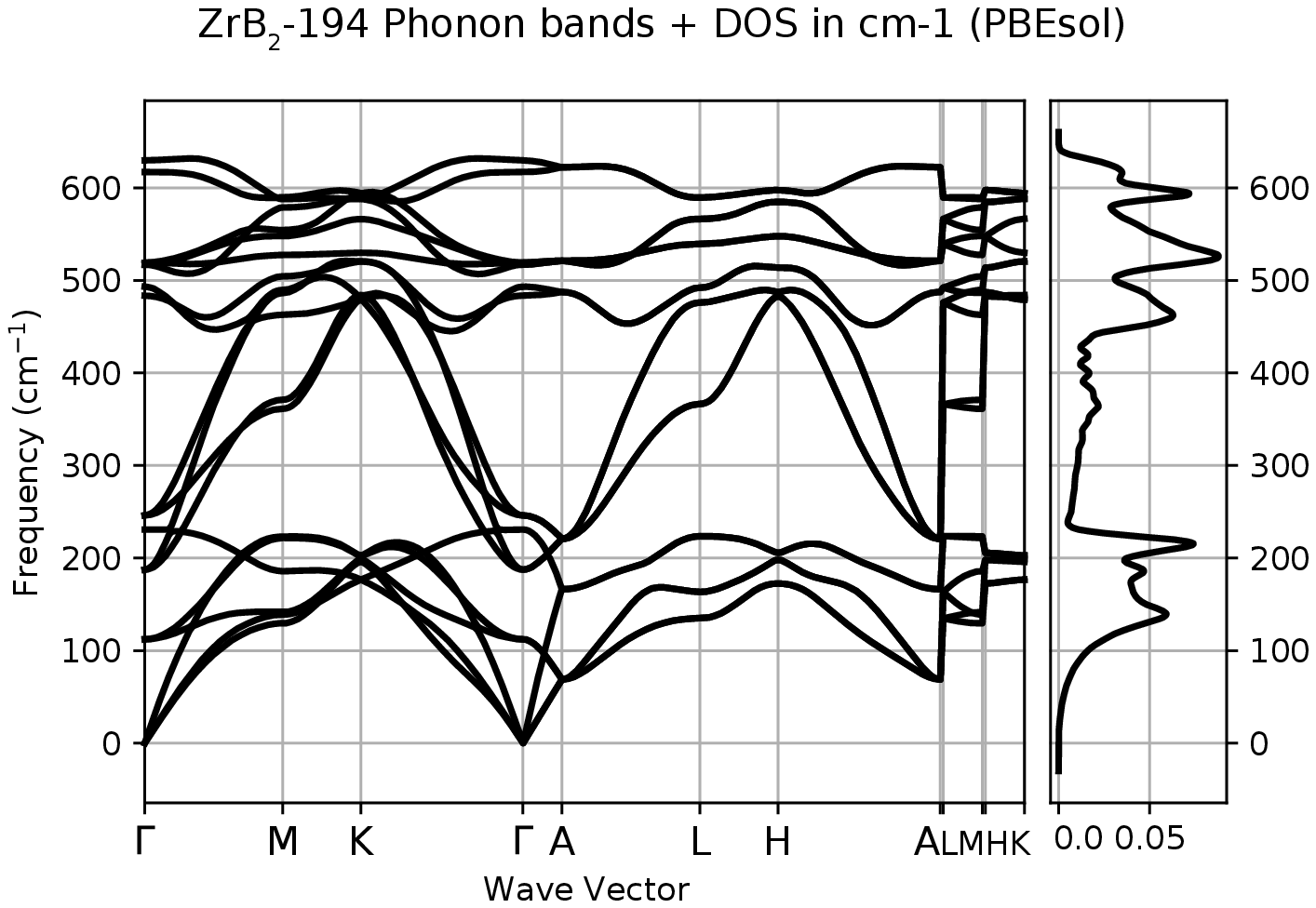} 
	\end{tabular}
	\caption{ZrB$_2$(\textit{hP6}-P6$_3$/mmc-space group, no.194)-Phonon band structure and DOS.}
	\label{fig:194PhononDispersion}
\end{figure}

\begin{figure}[H] 
	\centering
	\begin{tabular}{c}
		\includegraphics[width=0.98\linewidth]{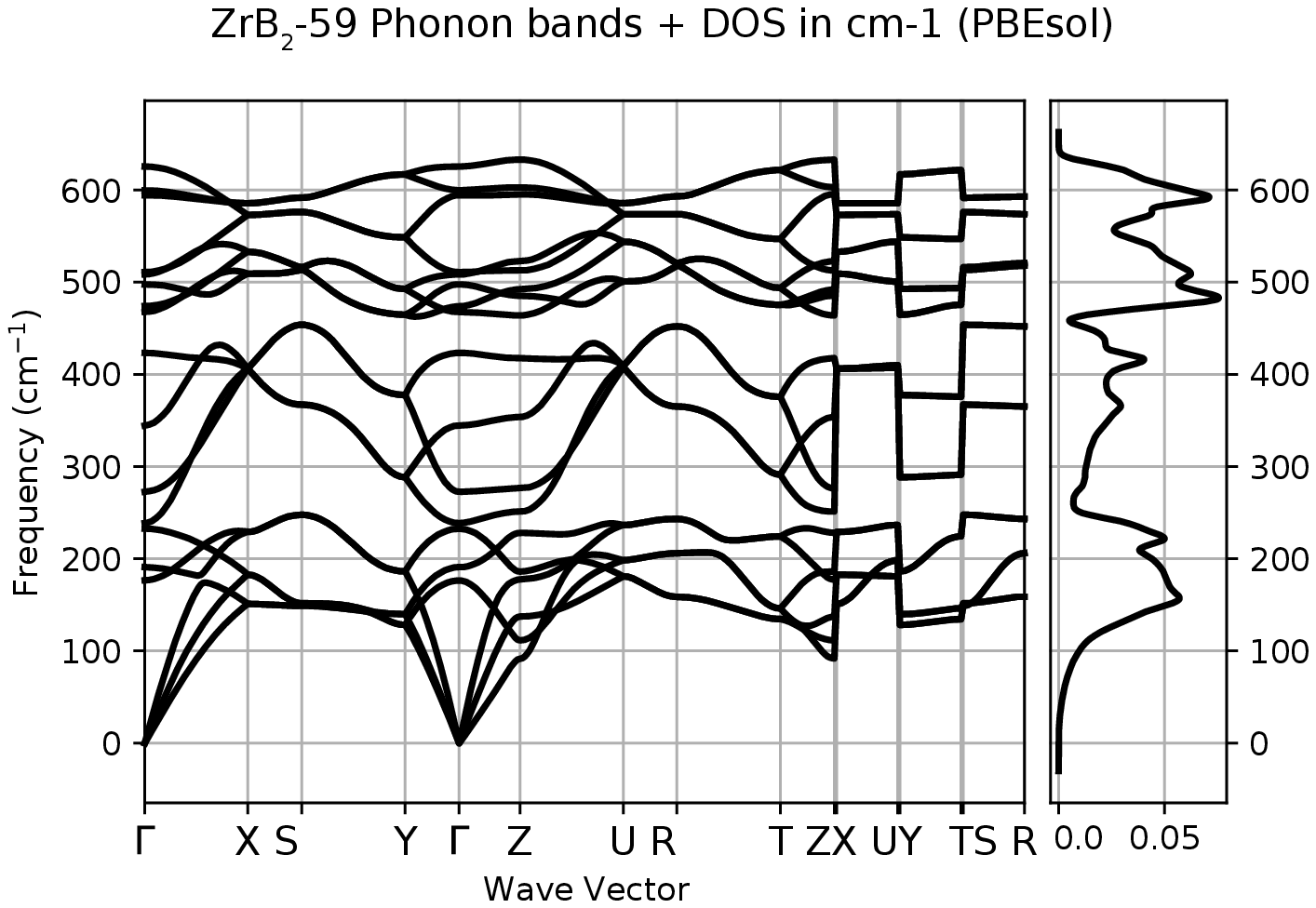} 
	\end{tabular}
	\caption{ZrB$_2$(\textit{oP6}-P6/mmm-space group, no.59)-Phonon band structure and DOS.}
	\label{fig:59PhononDispersion}
\end{figure}

\begin{figure}[H] 
	\centering
	\begin{tabular}{c}
		\includegraphics[width=0.90\linewidth]{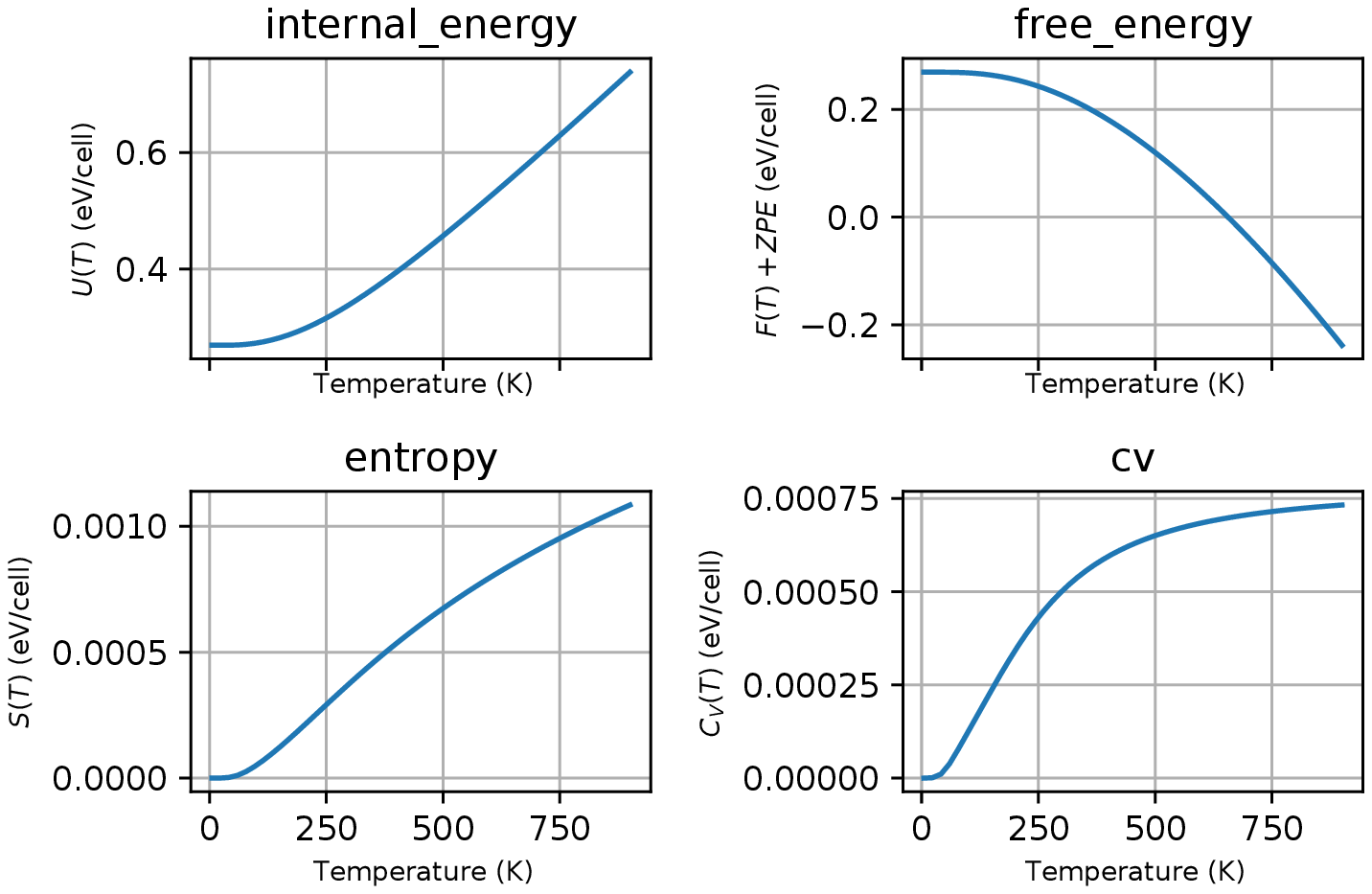} 
	\end{tabular}
	\caption{ZrB$_2$(\textit{hP3}-P6/mmm-space group, no.191)-Thermodynamic properties: internal energy, free energy, entropy and constant-volume specific heat.}
	\label{fig:191thermodynamic_properties}
\end{figure}

\begin{figure}[H] 
	\centering
	\begin{tabular}{c}
		\includegraphics[width=0.90\linewidth]{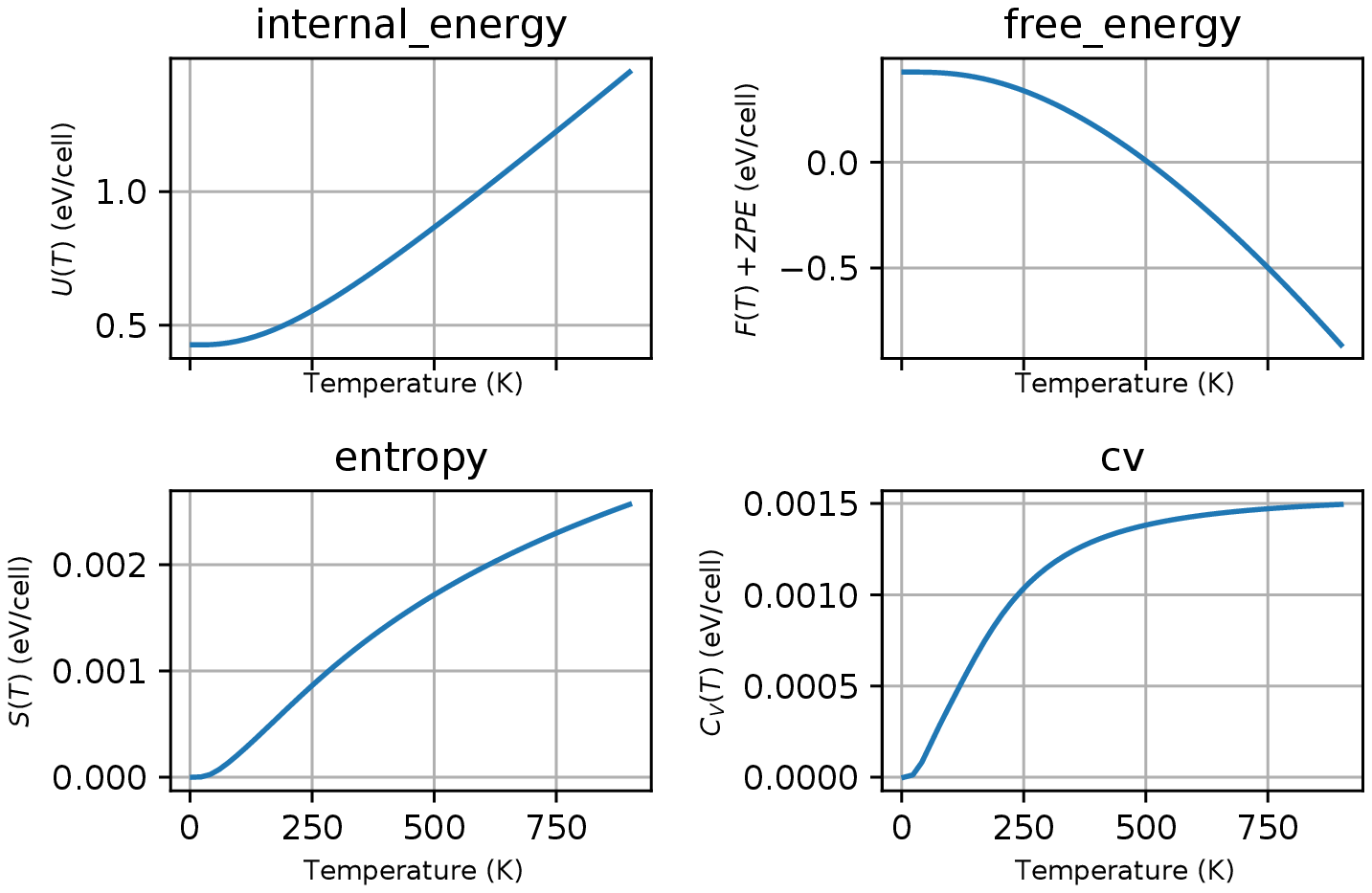} 
	\end{tabular}
	\caption{ZrB$_2$(\textit{hP6}-P6$_3$/mmc-space group, no.194)-Thermodynamic properties: internal energy, free energy, entropy and constant-volume specific heat.}
	\label{fig:194thermodynamic_properties}
\end{figure}

\begin{figure}[H] 
	\centering
	\begin{tabular}{c}
		\includegraphics[width=0.90\linewidth]{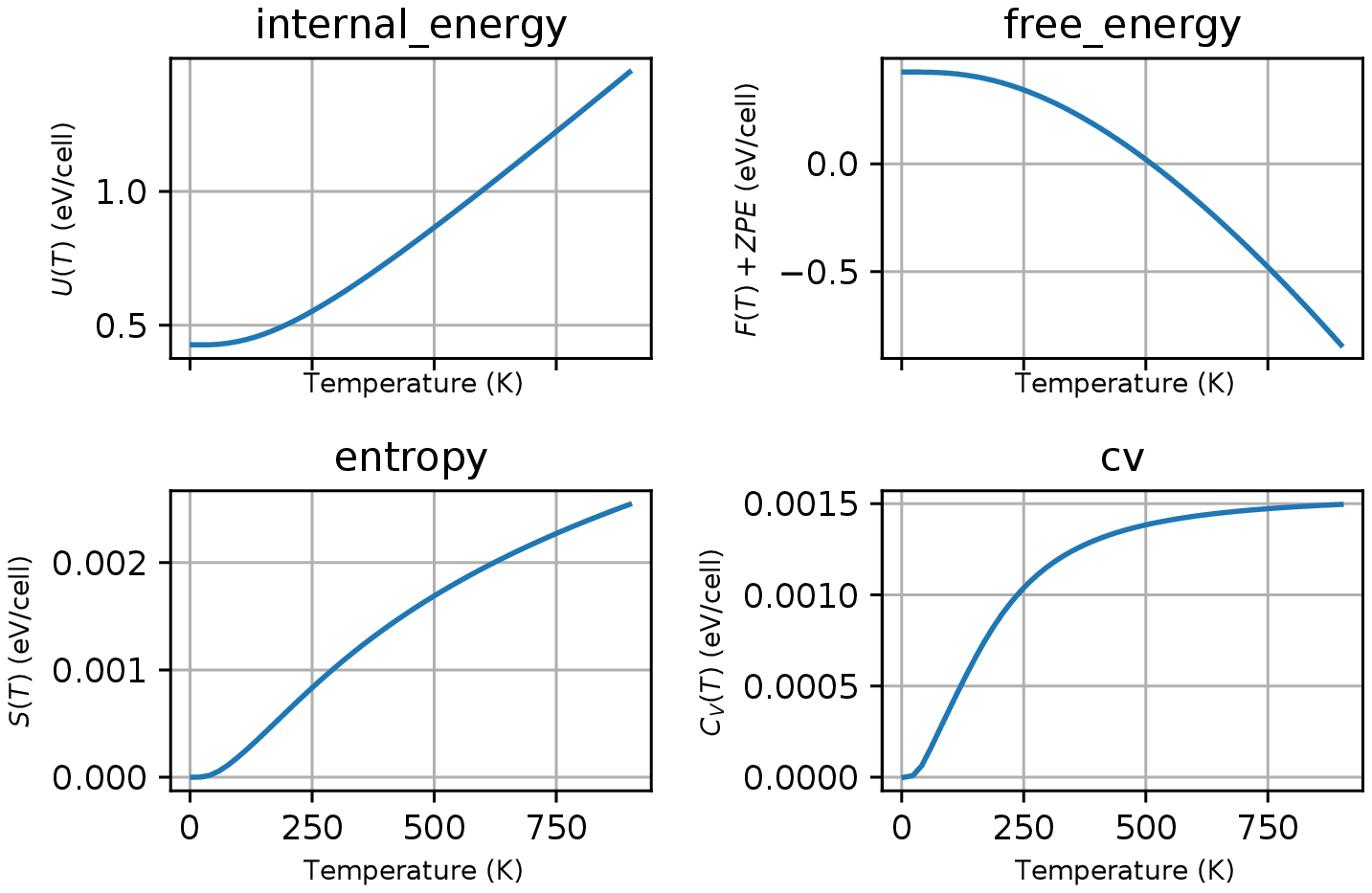} 
	\end{tabular}
	\caption{ZrB$_2$(\textit{oP6}-P6/mmm-space group, no.59)-Thermodynamic properties: internal energy, free energy, entropy and constant-volume specific heat.}
	\label{fig:59thermodynamic_properties}
\end{figure}

Phonon dispersion curves along the high symmetry q-points \cite{HINUMA2017140} and  phonon densities of states (DOS) calculated with the use of PBEsol exchange-correlation (XC) functional for known \textit{hP3}-P6/mmm phase, Fig.\ref{fig:ZrB2Basiccells} a), are depicted in Fig.\ref{fig:191PhononDispersion}, for the phase \textit{hP6}-P6$_3$/mmc, Fig.\ref{fig:ZrB2Basiccells} b), in Fig.\ref{fig:194PhononDispersion} and for the phase \textit{oP6}-Pmmn, Fig.\ref{fig:ZrB2Basiccells} c), in Fig.\ref{fig:59PhononDispersion}, respectively. Phonon results for all exchange-correlation (XC) functionals  are stored in {\ref{app:Appendix}. Supplementary data}. Analysis of the calculated curves allows us to state that phonon modes everywhere have positive frequencies and the new ZrB$_2$ phases are not only mechanically but also dynamically stable.
The estimated acoustic Debye temperature  $\Theta_D$ for the two new proposed ZrB$_2$ phases is about 760\,K and is about 200\,K lower than that for the known \textit{hP3}-P6/mmm phase and it is consistent with the mechanical properties, see Tab.\ref{tab:ZrB2Results}.
Results for thermodynamic properties up to 760\,K for the three zirconium diboride polymorphs calculated with the use of PBEsol exchange-correlation (XC) functional, i.e. {phonon internal energy}, {free energy}, {entropy}, {constant volume heat capacity} are depicted in Figs.\ref{fig:191thermodynamic_properties}, \ref{fig:194thermodynamic_properties}, \ref{fig:59thermodynamic_properties} and it can be seen that are very similar for the two new proposed ZrB$_2$ polymorphs.
This additional fact suggests again that the new phases are comparably thermodynamically stable up to Debye temperature  $\Theta_D$, but less stable than the known \textit{hP3}-P6/mmm phase.

\section{Conclusions}
\label{sec:Concl}

In the present paper extensive analysis of two new hypothetical and one previously known zirconium diboride (ZrB$_2$) polymorphs within the framework of density functional theory from the structural, mechanical and thermodynamic properties point of view was performed. We can conclude that:
\begin{itemize}
\item two new hypothetical zirconium diboride (ZrB$_2$) polymorphs: (\textit{hP6}-P6$_3$/mmc-space group, no.194) and (\textit{oP6}-Pmmn-space group, no.59) are mechanically and dynamically stable,
\item  these phases are comparably thermodynamically stable but less stable than the known \textit{hP3}-P6/mmm phase,  
\item \textit{hP6}-P6$_3$/mmc phase is ductile and \textit{oP6}-Pmmn phase is intermediate between brittle and ductile,
\item both new phases have a lower hardness than the known \textit{hP3}-P6/mmm phase. 
\end{itemize}
Many results in the paper, particularly relating to new hypothetical zirconium diboride (ZrB$_2$) polymorphs, are the first to be notified and we trust will be confirmed by other studies.

\section*{ACKNOWLEDGMENTS}
This work was supported by the National Science Centre (NCN -- Poland)  Research Project: UMO-2017/25/B/ST8/01789.
Additional assistance was granted through the computing cluster GRAFEN at Biocentrum Ochota, the Interdisciplinary Centre for Mathematical and Computational Modelling of Warsaw University (ICM UW) and Pozna\'n Supercomputing and Networking Center (PSNC). 

\appendix
\section{Supplementary data} 
\label{app:Appendix}

Supplementary data to this article can be found online at \href{run:./SupplementaryInformationCombined.pdf}{Supplementary Information}.


\bibliography{References}

\end{document}